\documentclass[aip,reprint]{revtex4-2}
\usepackage[T1]{fontenc}
\usepackage[utf8]{inputenc}
\usepackage{graphicx}
\usepackage{upgreek}
\usepackage{color}
\usepackage[
colorlinks=true,
linkcolor=blue,
citecolor=blue,
urlcolor = blue]{hyperref}

\draft % marks overfull lines with a black rule on the right

\begin{document}

%\preprint{}

\title{Single-spin scanning magnetic microscopy with radial basis function reconstruction algorithm}



\author{Cheng-Jie Wang}
\affiliation{Hefei National Laboratory for Physical Sciences at the Microscale and Department of Modern Physics, University of Science and Technology of China, Hefei~230026, China}
\affiliation{CAS Key Laboratory of Microscale Magnetic Resonance, University of Science and Technology of China, Hefei~230026, China}
\affiliation{Synergetic Innovation Center of Quantum Information and Quantum Physics,  University of Science and Technology of China, Hefei~230026, China}
\affiliation{Department of Physics, University of Science and Technology of China, Hefei~230026, China}

\author{Rui Li}
\affiliation{Hefei National Laboratory for Physical Sciences at the Microscale and Department of Modern Physics, University of Science and Technology of China, Hefei~230026, China}
\affiliation{CAS Key Laboratory of Microscale Magnetic Resonance, University of Science and Technology of China, Hefei~230026, China}
\affiliation{Synergetic Innovation Center of Quantum Information and Quantum Physics, University of Science and Technology of China, Hefei~230026, China}

\author{Bei Ding}
\affiliation{Beijing National Laboratory for Condensed Matter Physics and Institute of Physics, Chinese Academy of Sciences, Beijing 100190, China}

\author{Pengfei Wang}
\email{wpf@ustc.edu.cn}
\affiliation{Hefei National Laboratory for Physical Sciences at the Microscale  and Department of Modern Physics, University of Science and Technology of China, Hefei~230026, China}
\affiliation{CAS Key Laboratory of Microscale Magnetic Resonance, University of Science and Technology of China, Hefei~230026, China}
\affiliation{Synergetic Innovation Center of Quantum Information and Quantum Physics, University of Science and Technology of China, Hefei~230026, China}

\author{Wenhong Wang}
\affiliation{Beijing National Laboratory for Condensed Matter Physics and Institute of Physics, Chinese Academy of Sciences, Beijing 100190, China}

\author{Mengqi Wang}
\affiliation{Hefei National Laboratory for Physical Sciences at the Microscale and Department of Modern Physics, University of Science and Technology of China, Hefei~230026, China}
\affiliation{CAS Key Laboratory of Microscale Magnetic Resonance, University of Science and Technology of China, Hefei~230026, China}
\affiliation{Synergetic Innovation Center of Quantum Information and Quantum Physics, University of Science and Technology of China, Hefei~230026, China}

\author{Maosen Guo}
\affiliation{Hefei National Laboratory for Physical Sciences at the Microscale and Department of Modern Physics, University of Science and Technology of China, Hefei~230026, China}
\affiliation{CAS Key Laboratory of Microscale Magnetic Resonance, University of Science and Technology of China, Hefei~230026, China}
\affiliation{Synergetic Innovation Center of Quantum Information and Quantum Physics, University of Science and Technology of China, Hefei~230026, China}

\author{Chang-Kui Duan}
\email{ckduan@ustc.edu.cn}
\affiliation{Hefei National Laboratory for Physical Sciences at the Microscale and Department of Modern Physics, University of Science and Technology of China, Hefei~230026, China}
\affiliation{CAS Key Laboratory of Microscale Magnetic Resonance, University of Science and Technology of China, Hefei~230026, China}
\affiliation{Synergetic Innovation Center of Quantum Information and Quantum Physics, University of Science and Technology of China, Hefei~230026, China}
\affiliation{Department of Physics, University of Science and Technology of China, Hefei~230026, China}

\author{Fazhan Shi}
\affiliation{Hefei National Laboratory for Physical Sciences at the Microscale and Department of Modern Physics, University of Science and Technology of China, Hefei~230026, China}
\affiliation{CAS Key Laboratory of Microscale Magnetic Resonance, University of Science and Technology of China, Hefei~230026, China}
\affiliation{Synergetic Innovation Center of Quantum Information and Quantum Physics, University of Science and Technology of China, Hefei~230026, China}

\author{Jiangfeng Du}
\email{djf@ustc.edu.cn}
\affiliation{Hefei National Laboratory for Physical Sciences at the Microscale and Department of Modern Physics, University of Science and Technology of China, Hefei~230026, China}
\affiliation{CAS Key Laboratory of Microscale Magnetic Resonance, University of Science and Technology of China, Hefei~230026, China}
\affiliation{Synergetic Innovation Center of Quantum Information and Quantum Physics, University of Science and Technology of China, Hefei~230026, China}	


\date{\today}

\begin{abstract}
Exotic magnetic structures, such as magnetic skyrmions and domain walls, are becoming more important in nitrogen-vacancy center scanning magnetometry. However, a systematic imaging approach to mapping stray fields with fluctuation of several milliteslas generated by such structures is not yet available. Here we present a scheme to image a millitesla magnetic field by tracking the magnetic resonance frequency, which can record multiple contour lines for a magnetic field. The radial basis function algorithm is employed to reconstruct the magnetic field from the contour lines. Simulations with shot noise quantitatively confirm the high quality of the reconstruction algorithm. The method was validated by imaging the stray field of a frustrated magnet. Our scheme had a maximum detectable magnetic field gradient of 0.86 mT per pixel, which enables the efficient imaging of millitesla magnetic fields.
\end{abstract}

\maketitle %\maketitle must follow title, authors, abstract


Imaging the magnetic field generated by spins and currents is a powerful method for studying materials and devices. In recent years, a scanning magnetometer based on the nitrogen-vacancy color center (NV center) in diamond is emerging. It enables magnetic field imaging with high sensitivity and nanoscale spatial resolution. \cite{taylor_high-sensitivity_2008,casola_probing_2018} Measuring the Zeeman splitting of the energy levels of NV centers can be used to determine the magnetic field. Moreover, quantum interference schemes can be applied to improve the sensitivity. \cite{taylor_high-sensitivity_2008,maze_nanoscale_2008} After the first experiment that demonstrated potential nanoscale imaging magnetometry, \cite{balasubramanian_nanoscale_2008} NV center-based microscopy was recently adopted to study the non-collinear antiferromagnetic order, \cite{gross_real-space_2017} magnetic skyrmions, \cite{dovzhenko_magnetostatic_2018,gross_skyrmion_2018,akhtar_current-induced_2019} and magnetism in two-dimensional materials. \cite{thiel_probing_2019}

The range of the fluctuations of the stray fields generated by many magnetic structures is usually up to several milliteslas. \cite{tetienne_nature_2015,dovzhenko_magnetostatic_2018,yu_room-temperature_2018,jenkins_single-spin_2019,velez_high-speed_2019} Although a magnetic field with fluctuations of less than 1 mT can easily be measured using an optically detected magnetic resonance (ODMR) spectrum, magnetic fields with a large-fluctuation range cannot be measured because the fluctuations are beyond the range covered by the linewidth of a spectrum. To image a magnetic field with a large fluctuation range, either expanding the microwave (MW) frequency sweeping range or adjusting the range during scanning is required. However, large-range sweeping is time-consuming and it can be laborious to adjust a sweeping range with noise. Hence, an efficient scanning scheme that can map a magnetic field with fluctuations of several milliteslas is not available. \citet{haberle_high-dynamic-range_2013} optimized the control pulse to achieve imaging with a high dynamic range. However, the reconstruction of a magnetic field could be impossible if the change of the magnetic field between adjacent pixels is larger than the frequency gap of the grating pulse. This ability can be defined by the maximum detectable derivative of the magnetic fields with respect to space and hence, the acutance of the imaging method. Although the lock-in technology \cite{schoenfeld_real_2011} is also constrained by the linewidth, tracking the fluctuations of magnetic fields can result in high-acutance and high-dynamic-range imaging. However, a reconstruction algorithm is still required.

In this work, we develop and demonstrate a magnetic field tracking method and a reconstruction algorithm that enables the efficient and robust imaging of millitesla  magnetic fields. The tracking was accomplished by adjusting the MW frequencies based on the photoluminescence (PL) for three frequencies. The algorithm utilized radial basis functions (RBFs) to reconstruct the magnetic fields from the magnetic resonance fringe image.

\begin{figure}
	\includegraphics[scale=0.6]{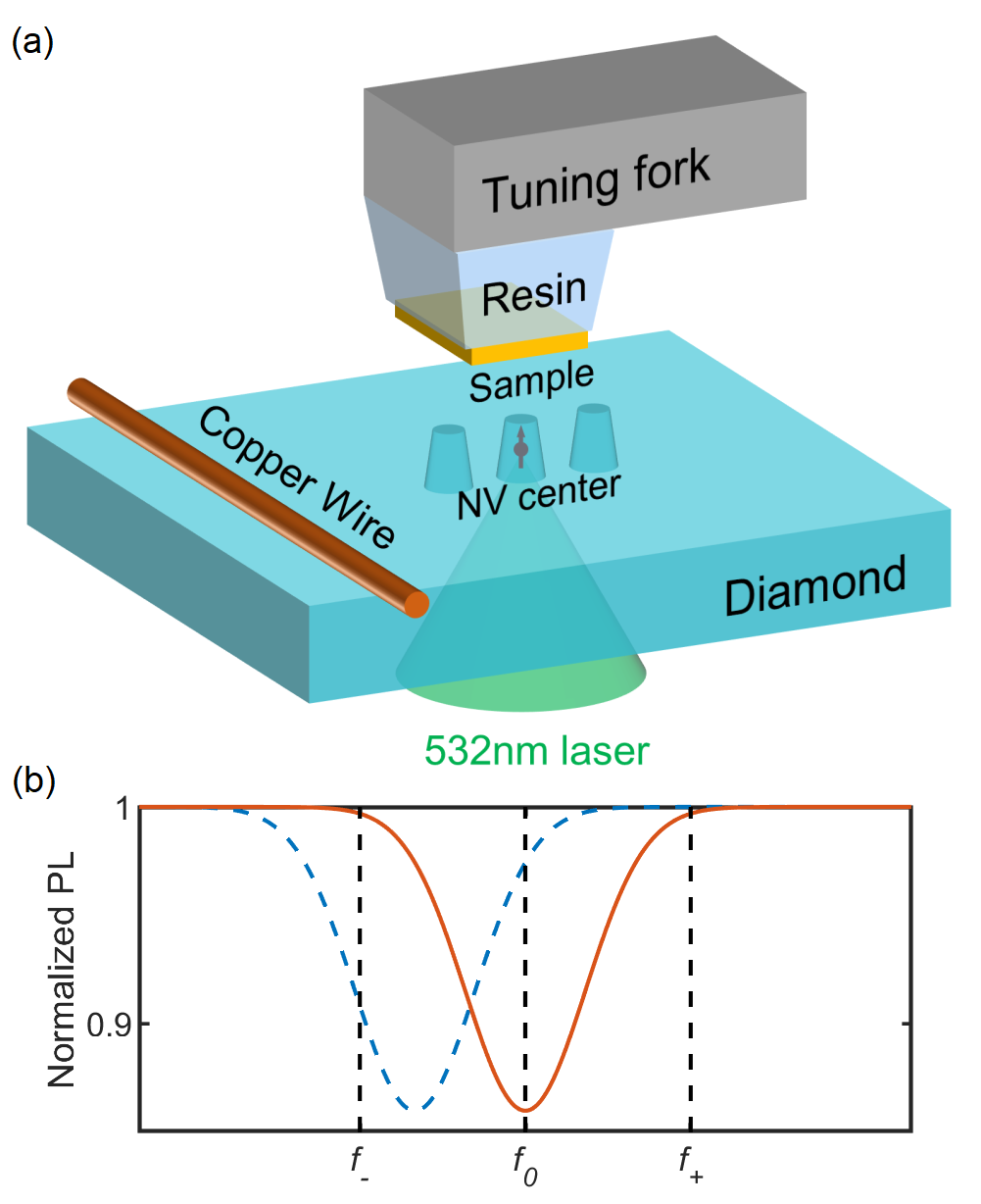}
	\caption{(a) Sketch of the measurement configuration. The sample is attached with polydimethylsiloxane (PDMS) to a resin platform, which is fixed onto a tuning fork. The sample is carried by the tuning fork and scanned over an NV center at a height of hundreds of nanometers. The NV center is in a pillar on the surface of a diamond bulk. The copper wire acts as an antenna to deliver MWs. (b) Illustration of the dynamic adjustment of MW frequencies in tracking the resonance frequency. Three vertical dashed lines from left to right are three excitation frequencies $ f_- $, $ f_0 $ and $ f_+ $, sequentially. The red solid line is the ODMR spectrum measured for the first pixel. The blue dashed line plots the case in which the down-shift of the MW frequencies is triggered.}\label{fig:principle}
\end{figure}

Recording the PL for three excitation frequencies is illustrated in Fig.~\ref{fig:principle}(b) (see the \href{}{supplementary material} for a flowchart). At first, the resonance frequency was determined with a full ODMR spectrum for the first pixel and used as one excitation frequency, denoted as $ f_0 $. The other two excitation frequencies were $ f_\pm=f_0\pm\delta $, where $ \delta $ is a fixed offset. $ \delta $ can be assigned as the full width at half maximum (FWHM) of the spectrum and adjusted according to the estimated magnetic field gradient. The PL was recorded while an MW field was applied with a fixed frequency, as in the iso-magnetic field method.\cite{balasubramanian_nanoscale_2008,rondin_nanoscale_2012,rondin_stray-field_2013} By contrast, we used three excitation frequencies $ f_0 $, $ f_- $, and $ f_+ $ for each pixel and so the PL was $ C_0 $, $ C_- $, and $ C_+ $, respectively. The $ C_0 $ signal was used to form the magnetic resonance fringe image, and $ C_- $ and $ C_+ $ were used to detect any changes of the resonance frequency from $ f_0 $ [Fig.~\ref{fig:principle}(b)]. $C_0$ was compared with $\min\{C_-,C_+\}$ and then all three frequencies ($ f_0 $ and $f_\pm$) were reduced (increased) by $\delta$ for the next pixel if $ C_- $ ($ C_+ $) was less than $ C_0 $. A threshold $ k $ was introduced to reduce the effect of PL fluctuations and other noise. The decision to change frequencies was made by comparing the ratio of $ C_0 $ to $ C_\pm $ with $ k $ instead of comparing $ C_0 $ with $C_\pm$. There is a tradeoff since $ k $ increases the robustness against noise at the cost of adding a delay in tracking the resonance frequency.

Through the above procedure, three maps of $ C_0 $, $ C_- $, and $ C_+ $ and one map of $ f_0 $ were recorded. The image of the resonance fringes, i.e., the normalized PL, was calculated for each pixel and used to estimate the ODMR contrast:
\begin{equation}
S=\frac{C_0}{\max\{C_{-},C_{+}\}}.
\end{equation}
Here, the maximum of $ C_- $ and $ C_+ $ was used as a reference to reduce the influence of PL fluctuations. This is because $ f_- $ or $ f_+ $ could be in the range of the spectrum, and thus, $ C_- $ or $ C_+ $ could be smaller than the reference PL [dashed line in Fig.~\ref{fig:principle}(b)]. In the post-processing, to increase the number of pixels with moderate contrast, we exchanged the value of $ C_0 $ with the value of $ C_- $ ($ C_+ $) for some pixels where the MW frequency down-shift (up-shift) was triggered during scanning, and changed the value of $ f_0 $ correspondingly.

The RBF algorithm was employed to reconstruct the magnetic field from the contour lines. RBFs are widely used to reconstruct surfaces with scattered data because they are easy to calculate and have good reconstruction quality. \cite{donato_approximate_2002,dinh_reconstructing_2002,cakmakci_application_2008} One desirable property of an RBF is its resistance against measurement noise since it takes the smoothness of the reconstructed surface as a constraint. Also, the meshless RBF method can discard invalid data. 

The RBF algorithm utilizes the sum of RBFs to approximate target data. One commonly used basis function is the thin plate spline (TPS) because it is easy to estimate its smoothness (also called the bending energy). Let $ S_i $ denote the target value at $ (x_i, y_i) $ in a map, where $ i =1, 2, \dots, n $. The TPS function, the approximation of the resonance frequency in this work, has the form:
\begin{equation}\label{eq:TPS}
f(x, y)=a_{1}+a_{2} x+a_{3} y+\sum_{i=1}^{n} b_{i} \phi\left(\left\|\left(x_{i}, y_{i}\right)-(x, y)\right\|\right),
\end{equation}
where $\left \|\cdot \right\| $ denotes the usual Euclidean norm and $ a_1 $, $ a_2 $, $ a_3 $, and $ b_i $ are coefficients to be determined. First, $ a_1 $, $ a_2 $, and $ a_3 $ can be determined by linear fitting. The TPS corresponds to the radial basis kernel $\phi(r)=r^2\log{r}$.

The smoothness or bending energy is calculated with the square integral of the second derivatives of $ f(x,y) $ instead of the curvature. For $f(x,y)$ to have square integrable second derivatives, the following conditions must be satisfied:\cite{donato_approximate_2002}
\begin{equation}
\sum_{i=1}^{n} b_{i}=\sum_{i=1}^{n} b_{i} x_{i}=\sum_{i=1}^{n} b_{i} y_{i}=0.
\end{equation}

Therefore, the fitting was accomplished by minimizing the energy function:
\begin{equation}\label{eq:energy-function}
E(f)=\sum_{i=1}^{n}\left(S_{i}-g\left(f\left(x_{i}, y_{i}\right)-{f_0}_i\right)\right)^{2}+\lambda \sum_{i, j=1}^{n} b_{i} K_{i j} b_{j},
\end{equation}
where $ K_{ij}=\phi(\left \| (x_i,y_i )-(x_j,y_j )\right\| ) $. $ \lambda $ is a positive scalar that controls the smoothness (the second term). It is proportional to the square integral of the second derivatives of $f(x,y)$. The value of $ \lambda $ can be assigned from a noise distribution. The function $ g(x) $ is the line-shape function of the ODMR spectrum (a Gaussian function in this work). The parameters can be fitted from the spectrum measured for the first pixel.

We introduced a weight $ w_i=3-2.5S_i $ to rank the data and guide the search. The weight was constructed as a monotonically decreasing function of $ S_i $ to rank the maximum contrast above zero contrast. For convenience, we chose a linear function. The parameters were determined to ensure any data points with zero contrast weigh half of those with the maximum contrast for a typical contrast range of 0.8--1. Consequently, the new energy function to be minimized is
\begin{equation}
E(f)=\sum_{i=1}^{n} w_{i}^{2}\left(S_{i}-g\left(f\left(x_{i}, y_{i}\right)-f_{0_{i}}\right)\right)^{2}+\lambda \sum_{i, j=1}^{n} b_{i} K_{i j} b_{j}.
\end{equation}
This is an optimization problem and so, the coefficients can be computed with a quasi-Newton method. Thus, the map of resonance frequencies can be recovered using the TPS function [Eq.~(\ref{eq:TPS})] and the magnetic field can be calculated from the resonance frequency. \cite{taylor_high-sensitivity_2008} In addition, fewer coefficients could be used in the TPS function [Eq.~(\ref{eq:TPS})] to reduce the computation time for a slight reduction in the reconstruction quality. 

\begin{figure}
	\includegraphics{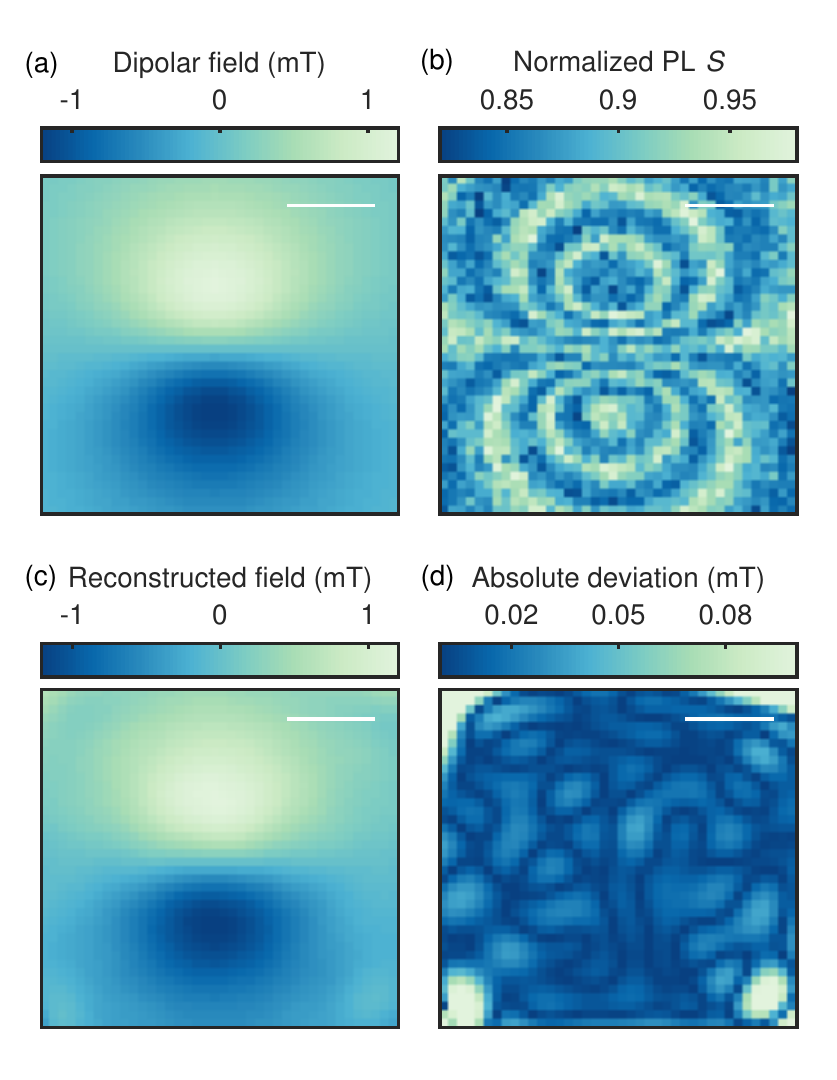}
	\caption{(a) The magnetic field for a simulated magnetic dipole. (b) Resonance fringe image calculated from (a) through the scanning procedure. Shot noise was introduced according to a Poisson distribution. The mean photon count was 5000. (c) The magnetic field reconstructed from (b). (d) Absolute deviations of (c) compared to (a). The scale bars are $250\,\mathrm{nm}$ in all panels.}\label{fig:simulation}
\end{figure}

To assess the reconstruction quality, a simulation was performed of a dipole-like field [Fig.~\ref{fig:simulation}(a)]. The parameters for the simulated dipolar field were chosen to produce a typical magnetic field in research. The other parameters in the simulation were the same as those used in the later experiments. As in the later experiments, a bias field of $ 5.5\,\mathrm{mT} $ aligned with the NV axis was applied to measure the magnetic field component along the NV axis from one resonance frequency of the two split energy levels. \cite{taylor_high-sensitivity_2008} In the simulated resonance fringe image, as shown in Fig.~\ref{fig:simulation}(b), the frequency used to excite each fringe was known exactly. Moreover, the reconstructed magnetic field [Fig.~\ref{fig:simulation}(c)] is consistent with the original dipolar field except for slight distortions in the corners. This is a boundary effect and could be suppressed. The absolute deviations between the reconstructed field and the original field in Fig.~\ref{fig:simulation}(d) quantitatively illustrate the reconstruction quality. As can be seen, the maximum deviation is about $ 0.04\,\mathrm{mT} $ except at the corners. The deviation is less than $ 0.03\,\mathrm{mT} $ in most pixels. Considering that the magnetic field ranges from $ 0.4\,\mathrm{mT} $ to $ 1.1\,\mathrm{mT} $ in most pixels, we concluded that the reconstruction field quantitatively agrees with the original field. Additional simulation experiments were performed with different photon counts  (see the \href{}{supplementary material}). These showed that the reconstruction algorithm is resilient to PL noise. As a result, the magnetic fields can always be reconstructed as long as there is no complete loss of tracking during scanning, which has a low probability because tracking loss is suppressed by the threshold $k$ and could probably be corrected for during the following scanning.

\begin{figure}
	\includegraphics{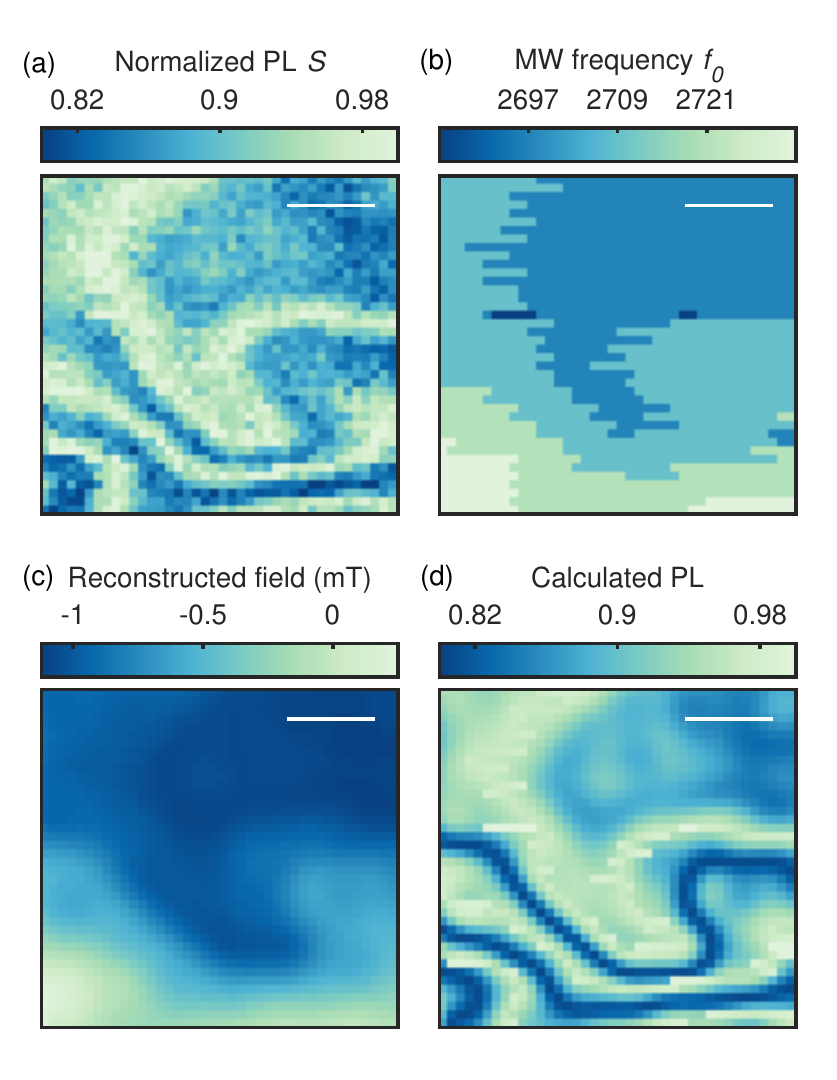}
	\caption{(a) Resonance fringe image obtained experimentally. (b) MW frequency map used to excite the NV center, i.e., the map of $ f_0 $. (c) Reconstructed magnetic field from (a). A bias field of $ 5.5\,\mathrm{mT} $ was subtracted. (d) Resonance fringe image calculated from (c). The scale bars are $ 250\,\mathrm{nm} $ in all panels.}\label{fig:exp1}
\end{figure}

Having assessed the magnetic field tracking and reconstruction algorithm, we applied this scheme to image the stray field of $ \mathrm{Fe}_3\mathrm{Sn}_2 $, which is a frustrated magnetic material with various spin textures. \cite{hou_observation_2017} Thus, its stray field is suitable for experimentally testing our scheme. The experimental setup is shown in Fig.~\ref{fig:principle}(a). The $ \mathrm{Fe}_3\mathrm{Sn}_2 $ thin film was attached to a resin platform. \cite{wang_nanoscale_2019} The atomic force microscope tip was a tuning fork, which carried the magnet to the NV centers. The resonance fringe image was acquired by scanning the sample over an NV center while recording the PL. We used software to track the resonant frequency during scanning in this work.

The experimental resonance fringe image is shown in Fig.~\ref{fig:exp1}(a). The parameters for the scanning procedure were: $ \delta=12\,\mathrm{MHz} $ and $ k=0.96 $. In the experiments, $ \lambda $ in the reconstruction algorithm was assigned to minimize the deviation between the calculated PL [Fig.~\ref{fig:exp1}(d)] and the normalized PL. $\lambda=1\times10^{-9}$ in this work can be used as a default value with the scanning scope normalized to $ 1\times1 $. The resonance fringe image is a sketch of the magnetic field in association with the map of $ f_0 $ [Fig.~\ref{fig:exp1}(b)]. There are no obvious distortions in the reconstructed field, as shown in Fig.~\ref{fig:exp1}(c). Furthermore, the resonance fringe image calculated from the reconstructed field is qualitatively in good agreement with the original one, which shows the accuracy of the reconstructed field.

We now discuss the performance of our scheme. First, the maximal sensitivity can be evaluated in the same way as an ODMR spectrum because the measurements in our scheme are based on the ODMR spectrum. The only difference is that the measurement duration is 1.5 times that for the ODMR spectrum because either $ C_- $ or $ C_+ $ is not utilized when calculating the normalized PL. Therefore, the sensitivity can be estimated from the PL rate and the contrast and linewidth of the ODMR spectrum. A sensitivity of $ 1.6\,\upmu \mathrm{T}/\sqrt{\mathrm{Hz}} $ was obtained in this work with the pulsed ODMR spectroscopy. \cite{dreau_avoiding_2011} Since $\delta$ is assigned as the FWHM and the maximum slope is in the range of the FWHM for either a Gaussian or Lorentz line shape, the PL can be measured at the maximum slope of the spectrum before changing MW frequencies.

Second, in a tracking approach, the acutance is determined by the difference in the magnetic field per step. In our scheme, the maximum detectable magnetic field gradient is twice $\delta$ per pixel beyond the usual limitation of the linewidth. Hence, $0.86\,\mathrm{mT}$ per pixel was achieved in our experiment. As a tracking method, it is easy to accomplish imaging with a high dynamic range as well. A magnetic field with a fluctuation range of $ 10\,\mathrm{mT} $ was obtained from the simulations (see the \href{}{supplementary material}), which is better by an order of magnitude compared with an ODMR spectrum of 10 bins and a bin size of 2--3\,MHz. It is also an improvement compared with the 2.2\,mT achieved by optimizing the control pulse. \cite{haberle_high-dynamic-range_2013} Note that the off-axis magnetic field could lead to a reduction in the ODMR contrast, though this effect can be handled well for an off-axis component of less than 10\,mT (see the \href{}{supplementary material}). However, any measurement based on an ODMR spectrum, including our scheme, would be invalid for larger off-axis magnetic fields. In this regime, the quenching method \cite{rondin_nanoscale_2012,tetienne_magnetic-field-dependent_2012,gross_skyrmion_2018} may be more appropriate. 

Third, our scanning scheme is pretty fast. The acquisition time for an image of 4096 pixels is about 5\,min, based on typical times for taking measurements (200 photons/ms and an integration time of 25\,ms/pixel for a frequency). Realizing an ODMR spectrum method would require 34\,min, including the acquisition of the PL reference. Given that the sensitivity is not improved, our scheme is an alternative when imaging millitesla magnetic fields efficiently, though with a compromise for precision.

In conclusion, we have presented a magnetic field tracking method that can map multiple contour lines of a magnetic field. It uses the TPS algorithm to reconstruct the magnetic field. Simulations show that the reconstructed field agrees well with the original magnetic field. The contour lines for the stray field of a frustrated magnet were measured experimentally and the reconstructed magnetic field was validated qualitatively. Our scheme has a maximum detectable magnetic field gradient beyond the linewidth of an ODMR spectrum without compromising speed or robustness. These advantages and other features of our approach make it more efficient for imaging millitesla magnetic fields, which will facilitate research based on NV magnetic microscopy.
\\
% If in two-column mode, this environment will change to single-column format so that long equations can be displayed. 
%\begin{widetext}
%$$\mbox{put long equation here}$$
%\end{widetext}

% Figures should be put into the text as floats. 
% Use the graphics or graphicx packages (distributed with LaTeX2e).
% Put the label that you will use with \ref{} command in the braces of the \label{} command.
%
% \begin{figure}
% \end{figure}

See the \href{}{supplementary material} for a flowchart of the tracking method, additional simulations, and a description of the preparation of the diamond sensor.
\\

The data that support the findings of this study are available from the corresponding author upon reasonable request.
\\

    This work was partially carried out at the USTC Center for Micro and Nanoscale Research and Fabrication. We acknowledge the financial support by the NNSFC (Grants No. 81788101, No. 11761131011, No. 61635012 and No. 11874338), the National Key R\&D Program of China (Grant No. 2018YFA0306600, No. 2018YFF01012500, No. 2017YFA0303202), the CAS (Grants No. GJJSTD20170001, No. QYZDY-SSW-SLH004, No. XDC07000000), and Anhui Initiative in Quantum Information Technologies (Grant No. AHY050000).

\bibliography{paper}

\end{document}

\title{Supplemental Material}
	\maketitle
	% Define block styles

	\begin{figure}
		\tikzstyle{startstop} = [rectangle, draw,  
		text width=5em, text centered, rounded corners, minimum height=3em]
		\tikzstyle{decision} = [diamond, draw, aspect=2,
		text width=9em, text badly centered, node distance=3cm, inner sep=0pt]
		\tikzstyle{process} = [rectangle, draw, 
		text width=10em, text centered, minimum height=3em]
		\tikzstyle{line} = [draw, -latex']
		\begin{tikzpicture}[node distance = 3cm, auto]
		% Place nodes
		\node[startstop](start){Start};
		\node[process,below of=start,node distance=2.5cm](counts){Record $ C_0$, $C_-$, $C_+ $};
		\node[decision,below of=counts ](compare0){$ C_-<C_+ $?};
		\node[decision,below of=compare0](compare1){$ kC_0 >C_- $?};
		\node[process,below of=compare1,node distance=3cm](decrease0){$ f_0=f_0-\delta $, $ f_\pm=f_\pm-\delta $};
		\node[decision,right of=compare1,node distance=6cm](compare2){$ kC_0 >C_+ $?};
		\node[process,below of=compare2,node distance=3cm](increase0){$ f_0=f_0+\delta $, $ f_\pm=f_\pm+\delta $ };
		\node[decision,below of=decrease0](finish){Finish scanning?};
		\node[process,left of=compare0, node distance=6cm](next){Move to next pixel};
		\node[startstop,below of=finish,node distance=3cm](end){End};
		% Draw edges
		\path[line](start)--(counts);
		%	\path[line](spectrum)--(f-+);
		%	\path[line](f-+)--(counts);
		\path[line](counts)--(compare0);
		\path[line](compare0)--node{yes}(compare1);
		\path[line](compare0)-|node{no}(compare2);
		\path[line](compare1)--node{yes}(decrease0);
		\path[line](compare2)--node{yes}(increase0);
		\path[line](decrease0)--(finish);
		\path[line](increase0)|-(finish);
		\path[line](finish)-|node{no}(next);
		\path[line](next)|-(counts);
		\path[line](finish)--node{yes}(end);
		\end{tikzpicture}
		\caption{A flowchart of magnetic field tracking during scanning.}
	\end{figure}

	\begin{figure}
		\includegraphics[scale=0.6]{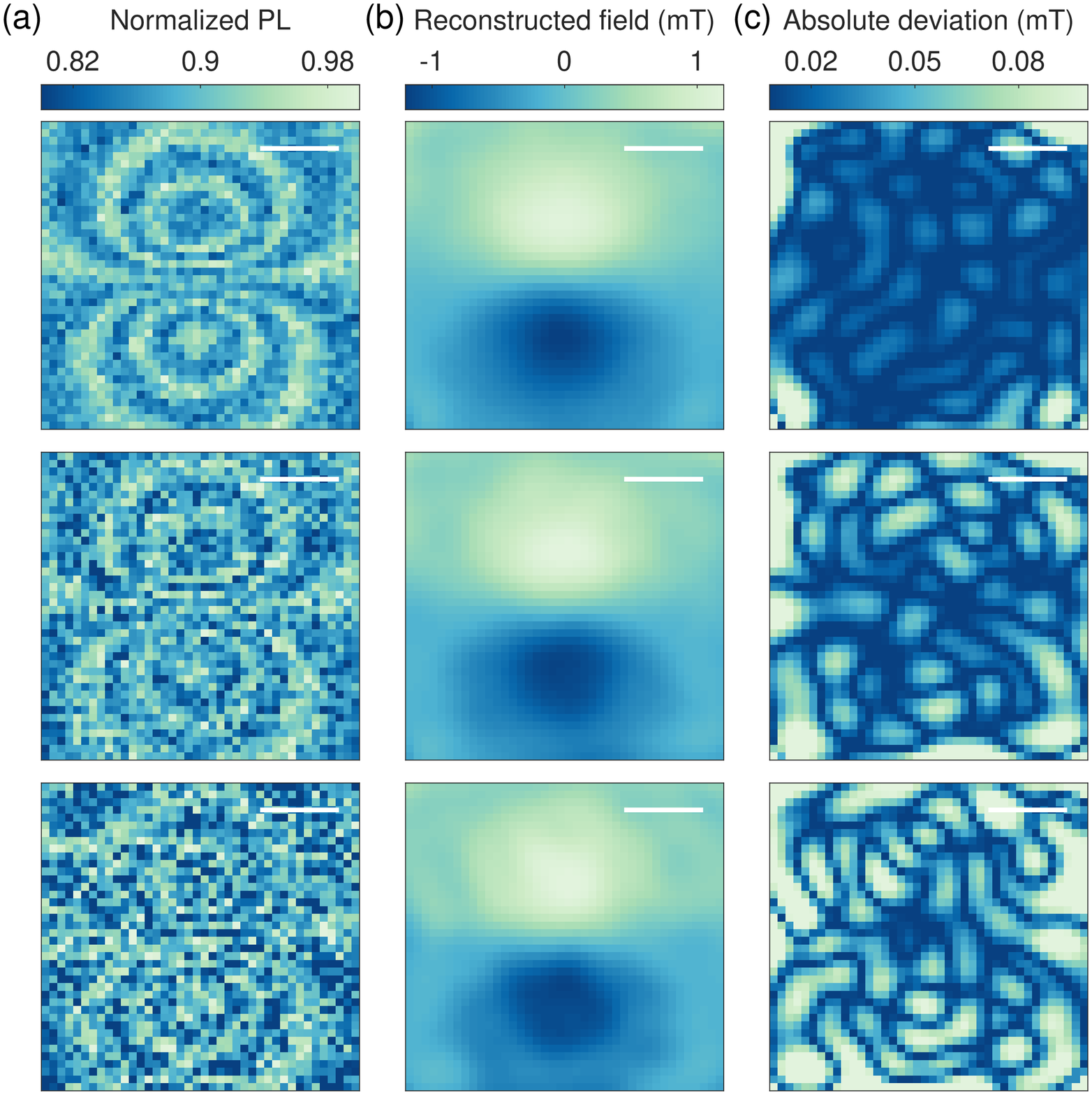}
		\caption{ Simulation experiments with different mean photon counts. (a) Magnetic resonance fringe image with mean photon counts of 3,000, 1,000 and 500 from top to bottom respectively. (b) Reconstructed magnetic field in experiments with mean photon counts of 3,000, 1,000 and 500 from top to bottom respectively. (c) Absolute deviation map in simulations with mean photon counts of 3,000, 1,000 and 500 from top to bottom respectively. All parameters are same as the simulation in the main text. The scale bar is $ 250~\mathrm{nm} $ for all panels.}
	\end{figure}

	\begin{figure}
	\includegraphics[scale=0.5]{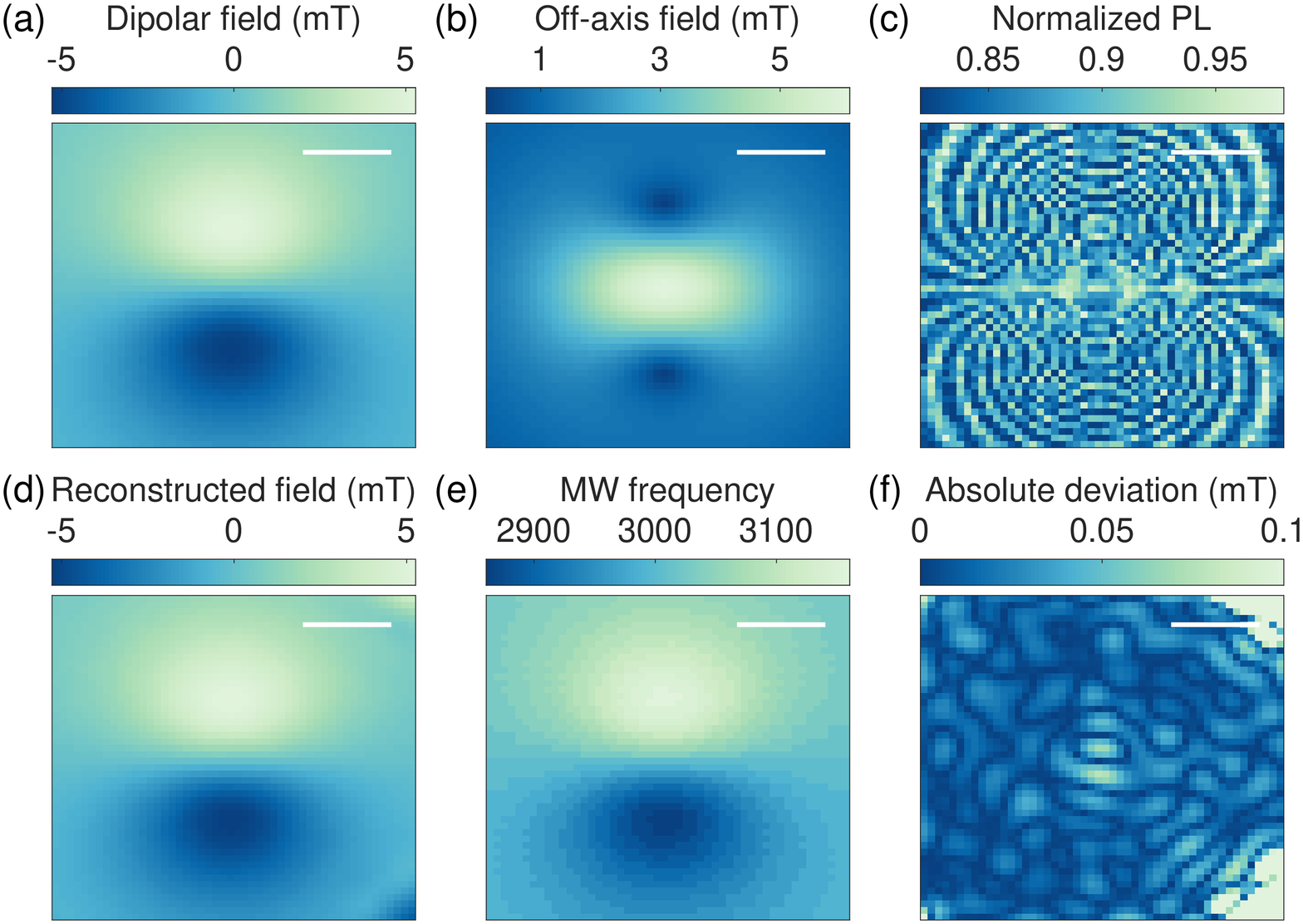}
	\caption{\label{fig:simulation} A simulation to image a magnetic field of 5 mT. (a) The on-axis component of a simulation magnetic dipolar field. (b) The off-axis component of the dipolar field. (c) Resonance fringe image calculated from (a) through the scanning procedure. Shot noise is introduced according to Poisson distribution and mean photon count is 5, 000. The reduction of spectrum contrast induced by the off-axis magnetic field is computed with a seven-level model \cite{tetienne_magnetic-field-dependent_2012} and parameters are from NV D in \cite{tetienne_magnetic-field-dependent_2012}. (d) The reconstructed magnetic field from (c). (e) The MW frequency map used to excite NV center, which roughly presents the magnetic field. (f) Absolute deviations of (d) from (a). The maximum deviation caused by the off-axis field is about 0.077 mT. The scale bar is $250~\mathrm{nm}$ for all panels. A bias field of 10 mT was aligned with the NV axis to retain the linear dependence of resonance frequencies on the magnetic field component projected on the NV axis. All parameters are same as the simulation in the main text except the ODMR contrast.}
\end{figure}
	\clearpage
\section*{Diamond sensor preparation}
The diamond used in our experiments is an electronic grade [111]-cut diamond grown by chemical vapor deposition so that the NV axis is aligned with the surface normal. The NV centers are formed by 37 keV $\mathrm{CN}^-$  ion implantation with a density of $1\times10^{11}/\mathrm{cm}^2$, which results in an estimation depth of about 30 nm. After annealing in vacuum at 800$^\circ\mathrm{C}$ for 4 hours and then in oxygen atmosphere at 580$^\circ\mathrm{ C}$ for 20 min, nanopillar grids with gaps of 10 $\upmu\mathrm{m}$ are fabricated on the diamond using electron beam lithography and reactive ion etching.